\documentclass[conference,a4paper]{IEEEtran}
\usepackage[utf8]{inputenc}
\usepackage{amsmath}
\usepackage{amsfonts}
\usepackage{amssymb}
\usepackage[english]{babel}
\usepackage{graphicx}
\usepackage{comment}
\usepackage{cite}
\usepackage{cmap}
\usepackage{setspace}
\usepackage{xcolor}
\usepackage{multicol}
\newtheorem{example}{Example}
\usepackage{tabularx,ragged2e,booktabs}
\newcolumntype{L}{>{\RaggedRight\arraybackslash}X} % ragged-right version of "X"

\begin{document}

\title{New Code-Based Cryptosystem with Arbitrary Error Vectors
\thanks{*Identify applicable funding agency here. If none, delete this.}
}
\author{\IEEEauthorblockN{Fedor Ivanov}
\IEEEauthorblockN{
National Research University\\
Higher School of Economics, \\
Institute for Information\\
Transmission Problems\\
Russian Academy of Sciences\\
Moscow, Russia \\
fivanov@hse.ru}
\and
\IEEEauthorblockN{Eugenii Krouk}
\IEEEauthorblockN{National Research University\\
Higher School of Economics, \\
Moscow, Russia \\
ekrouk@hse.ru}}

\maketitle
\begin{abstract}
McEliece cryptosystem represents a smart open key system based on the hardness of the decoding of an arbitrary linear code, which is believed to be able to resist the advent of quantum computers. But the original McEliece cryptosystem, based on Goppa codes, has just very limited interest in practice, partly because it requires a very large public key. In this paper we propose a new general way to reduce the public key size. Unlike most papers on reducing key length of the cryptosystem, where original Goppa codes are substituted by some other codes, we suggest a new method of key size reduction which is code-independent.     
\end{abstract}

\section{Introduction}
In 1976 W. Diffie and M. Hellman proposed the public-key cryptography concept  \cite{DH}. To construct the public-key cryptosystem one need to construct the one-way trap-door function. To achieve this, the hard computational problem should be selected, which nevertheless has simple solutions in some special cases. It is supposed that attacker (intruder) who desires to ''break'' the system should solve the hard problem (i.e. compute the correspondent plaintext from given ciphertext), while legitimate user, using his private key, obtains the simple special case of the hard problem and solves it for decryption.

However, to break the system one may not search for solution of the hard problem being used, but look for another function instead, providing decryption without knowledge of the private key. If such function would be computationally efficient, this leads to breaking the system without solution of initial hard problem. Excellent example of such attack on Merkle-Hellman cryptosystem \cite{MH} was given by A. Shamir \cite{Shamir}. An attempt of similar attack on code-based cryptosystems was made in \cite{SS}.

Code-based McEliece cryptosystem was proposed in 1978 \cite{RMC}. Task which is a basis of McEliece cryptosystem is a decoding of linear codes. It is well-known that the decoding of an arbitrary linear code is computationally difficult task but there are some classes of codes, say RS, BCH, LDPC codes that have polynomial-time decoding algorithms. The main idea underlying McEliece cryptosystem is to hide structured code with simple decoding algorithm (secret key) presenting it as a random code (open key) for which a simple decoder is unknown. 

Original McEliece cryptosystem gain no wide practical usage, and the most common-used public-key cryptosystem for today is RSA \cite{RSA}. However, RSA is based on intractability of integer factorization, for which the polynomial-time quantum algorithm was proposed by Shor \cite{Shor}. At the same time no effective quantum algorithm is known to break McEliece cryptosystem. 

The main drawback of McEliece cryptosystem is a very long public key. For example, in the original paper by McEliece open key has size $262000$ bits. 

There are a number of attempts to overcome this main disadvantage of code-based cryptosystem. The main of them have the same idea of substituting the original Goppa code that is used in McEliece cryptosystem to some other one with a specific structure that allows to reduce open key size. For instance in paper \cite{grs} Goppa codes are substituted by subfield subcodes of quasi-cyclic generalized Reed-Solomon codes. It allows to obtain cryptosystem with key size from $6000$ to $11000$ bits with security ranging from $2^{80}$ to $2^{107}$. In paper \cite{qcmdpc} quasi-cyclic moderate-density parity-check codes (QC-MDPC) were suggested to implement in McElece cryptosystem. It results in significant key size reduction up to $0.6$ KBytes that makes code-based cryptosystem practically feasible. Very similar instance of the cryptosistem is based on QC-LDPC codes and was proposed in \cite{ldpcm}. In paper \cite{fired} was shown how to adapt QC-MDPC based cryptosystem for such low-power devices as RFID.

The main drawback of substituting Goppa codes by either QC-LDPC or QC-MDPC codes is as follows: it is well-known that practical hard-decoding algorithms of QC-LDPC and QC-MDPC codes does not take into account the minimal distance of these codes. Thus for a given error vector $\bf e$ with fixed weight $t$ we can not guarantee that it will be decoded with zero probability of error. Moverover, practically used QC-LDPC or QC-MDPC codes usually have rather small minimal distance (of the order of tens for codes of rate $R=\frac12$ and a length of several thousand). 

In this paper we suggest another approach of key length reduction. It is a universal and can be applied for any $(n,k)$ code. The proposed cryptosystem is based on the approach that was suggested in \cite{fiEurc}, but has some significant improvements: it is stable for both syndrome-based and rank attacks and also can be constructed for any linear block code even for the cases when this code does not have simple polynomial-time decoding algorithm. 

\section{McEliece Original Cryptosystem}

Let us describe a concept of McEliece open key cryptosystem. 
 
Let us assume that plaintext is represented as a $q$-ary vector $\bf u$ of length $k$. Let us assume that we have some $(n,k)$ $t$-error correcting code $C$, represented by it's generator matrix $\bf G$ and there is a polynomial time decoder $\xi$ of $C$. If we denote by $\phi$ and $\phi^{-1}$ encryption and decryption procedures respectively, then code-based cryptosystem can be described as follows:
\begin{enumerate}
    \item \textit{Encryption:} calculate $\phi({\bf u})={\bf c}={\bf u}{\bf G}'+{\bf e}={\bf u}{\bf S}{\bf G}{\bf P}+{\bf e}$, where $\bf S$ some non-singular $k\times k$ random matrix over $\mathbb{F}_q$, $\bf P$ is $n\times n$ permutation matrix and $\bf e$ is a $q$-ary random vector: $w({\bf e})\leq t$, i. e. weight of $\bf e$ is not greater then code $C$ correction capability $t$.
    \item \textit{Decryption:} calculate $\phi^{-1}({\bf c})$ by the following steps:
    \begin{itemize}
        \item ${\bf c}{\bf P^{-1}}={\bf u}{\bf S}{\bf G}+{\bf e}{\bf P}^{-1}$, where $w({\bf e}{\bf P}^{-1})=w({\bf e}).$
        \item Decode ${\bf c}{\bf P^{-1}}$ applying decoding algorithm $\xi$ of code $C$: $\xi({\bf c}{\bf P^{-1}})={\bf u}{\bf S}$. After this step we obtain some linear combination of transmitted word $\bf u$
        \item Calculate ${\bf u} = {\bf u}{\bf S}{\bf S}^{-1}$.
        \end{itemize}
\end{enumerate}

In this case the open key in McEliece cryptosystem is $\left({\bf G}',\:t\right)$ and a secret one is a factorization $\left({\bf S},{\bf G},{\bf P}\right)$. Without knowledge of secret key intruder has to solve complex task of decoding an arbitrary code described by it's generator matrix ${\bf G}'$. This task is known to have an exponential complexity.

\section{Information Set Decoding}

An attack based on Information Set Decoding (ISD) is known as one of the best ones against McEliece cryptosystem. This type of attack was already mentioned in the initial security analysis of McEliece \cite{RMC} and further explored by Lee and Brickell \cite{LBR}. There are a different modifications of this decoding with the best known complexity of $\tilde {\mathcal {O}}(2^{0.054 n})$  \cite{May}. In this paper we describe the general concept of this decoding. 

The task of ISD decoder is to recover information vector $\bf u$ from ${\bf c}={\bf u}{\bf G}+{\bf e}$, where $\bf G$ is a generator matrix of $(n,k)$ code $C$ with minimal distance $d=2t+1$ and $wt({\bf e})\leq t$.

Let us denote by $\mathcal{J}$ some subset of $\{1,2,\ldots, n\}$ and by ${\bf G}_{\mathcal{J}}$ the matrix which contains only columns at indexes from $\mathcal{J}$ of $\bf G$. At the same manner by ${\bf e}_{\mathcal{J}}$ we denote $|\mathcal{J}|$-length vector with elements of $\bf e$ that are placed in indexes from $\mathcal{J}$.

In this case to obtain $\bf u$ one applies the following algorithm:
\begin{enumerate}
    \item Randomly choose $k$ indexes $\mathcal{J}\subset\{1,2,\ldots, n\}.$
    \item If there are no errors in ${\bf c}_{\mathcal{J}}$ then the following relationship holds: ${\bf c}_{\mathcal{J}}={\bf u}_{\mathcal{J}}{\bf G}_{\mathcal{J}}+{\bf e}_{\mathcal{J}}.$
    \item If $wt\left({\bf c}+{\bf c}_{\mathcal{J}}{\bf G}_{\mathcal{J}}^{-1}{\bf G}\right)=t$ then there is no error in ${\bf c}_{\mathcal{J}}$ and thus $wt({\bf e}_{\mathcal{J}})=0$. In this case ${\bf u}={\bf c}_{\mathcal{J}}{\bf G}_{\mathcal{J}}^{-1}$. Else return to Step 1.
\end{enumerate}
It is easy to notice that the probability $P_s$ of having no errors in the chosen $k$ indexes is
$$
P_s=\dfrac{\binom{n-t}{k}}{\binom{n}{k}}=\dfrac{\binom{n-k}{t}}{\binom{n}{t}}.
$$

Thus the average number of decoding attempts $\tau$ before finding free of errors information set is
$$
\tau=\frac1{P_s}=\dfrac{\binom{n}{t}}{\binom{n-k}{t}}.
$$

This value $\tau$ can be considered as a complexity of ISD decoding algorithm.

In the next section we will describe new cryptosystem and show that attack based on ISD decoding can not be applied to it. Thus in order to obtain the same security level, significantly shorter codes can be used in new system.

\section{New Code-Based Cryptosystem}

Let us first describe the main parts of suggested code-based cryptosystem. 

We consider some $(n,k)$ code $C$. We will denote a generating matrix of this code by $\bf G$ and any fixed information set of this code by $\mathcal{J}$. Let us also consider $n\times n$ $q$-ary matrix ${\bf G}_0$ that consists of $n$ arbitrary codewords of $C$:
$$
{\bf G}_0=
\begin{pmatrix}
{\bf c}_1\\
{\bf c}_2\\
\ldots\\
{\bf c}_n
\end{pmatrix},
$$
where ${\bf c}_i\in C$, $i=1..n$.

Let us denote by ${\bf M}$ and $\bf T$ some arbitrary and non-singular $n\times n$ matrices over $\mathbb{F}_q$. We also consider such $n \times n$ matrix $\bf Q$ that:

$$
{\bf QT}=\left( {\bf 0}_{\mathcal{J}}\:\: {\bf R}_{[n]\setminus \mathcal{J}}\right),
$$
where $\mathcal{J}$ is an informational set of code $C$, $[n]\setminus \mathcal{J}$ is the complement of set $\mathcal{J}$, $\bf 0$ is $n\times k$ zero matrix, $\bf R$ is $n \times (n-k)$ full-rank matrix. It means that matrix $\bf QT$ has zero columns in positions of information set $\mathcal{J}$ of $C$ and other $n-k$ columns are non-zero and forms full-rank matrix $\bf R$. 

The proposed cryptosystem is as follows:

\begin{itemize}
    \item \textit{Open key:} $\left({\bf G}',{\bf G}_2'\right)=\left({\bf G}{\bf M}, {\bf Q}({\bf G}_0+{\bf T}){\bf M}\right)$
    \item \textit{Private key:} $\left({\bf G},{\bf M},{\bf T},{\bf Q},{\bf G}_0,\mathcal{J}\right)$
\end{itemize}

Encryption/decryption are as follows:

\begin{enumerate}
    \item \textit{Encryption:} Let us consider that we have some plaintext $\bf u$ which is a $q$-ary vector of length $k$. The encrypted text is: $\phi({\bf u})={\bf c}={\bf u}{\bf G}'+{\bf e}{\bf G}'_2$, where $\bf e$ is a $q$-ary random vector of arbitrary weight.
    \item \textit{Decryption:} Calculate $\phi^{-1}({\bf c})$ by the following steps:
    \begin{itemize}
        \item ${\bf y}={\bf c}{\bf M^{-1}}={\bf u}{\bf G}+{\bf e}{\bf Q}({\bf G}_0+{\bf T})={\bf u}{\bf G}+{\bf e}{\bf Q}{\bf G}_0+{\bf e}{\bf Q}{\bf T}$. 
        \item Notice that ${\bf u}{\bf G}+{\bf e}{\bf Q}{\bf G}_0\in C$ and vector ${\bf e}{\bf Q}{\bf T}$ has non-zero positions only on $[n]\setminus \mathcal{J}$, i. e. information set $\mathcal{J}$ is free of errors. Thus:
        $$
        {\bf y}-{\bf y}_{\mathcal{J}}{\bf G}_{\mathcal{J}}^{-1}{\bf G}={\bf e}{\bf QT}
        $$
        \item Calculate ${\bf e}{\bf QT}{\bf T}^{-1}={\bf e}{\bf Q}$
        \item Calculate ${\bf e}{\bf Q}{\bf G}_0$
        \item Find $\bf u$ from ${\bf u}{\bf G}={\bf y}-{\bf e}{\bf Q}{\bf G}_0-{\bf e}{\bf Q}{\bf T}$
        \end{itemize}
\end{enumerate}

Let us show how to construct matrices $\bf Q$ and $\bf T$ such that $\bf QT$ has rank $n-k$ and has $k$ zero columns at positions $\mathcal{J}$.

Let $\bf T$ any non-singular matrix and ${\bf T}_{\mathcal{J}}$ is the matrix which contains only columns at indexes from $\mathcal{J}$ of $\bf T$. If we consider matrix ${\bf T}_{\mathcal{J}}^T$ as a generator matrix of some code, then there exist a matrix ${\bf X}_{\mathcal{J}}$ of size $(n-k)\times n$ such that ${\bf X}_{\mathcal{J}}{\bf T}_{\mathcal{J}}={\bf 0}$. Let us construct matrix ${\bf Q}={\bf L}{\bf X}_{\mathcal{J}}$, where $\bf L$ is $(n-k)\times n$ arbitrary full-rank matrix. It is obvious that matrix ${\bf QT}={\bf L}{\bf X}_{\mathcal{J}}{\bf T}$ has zero columns at positions $\mathcal{J}$ and other  columns form full-rank matrix.

\section{Attacks against proposed cryptosystem}

In this section we discuss some attacks on proposed cryptosystem: direct attacks, syndrome-based attacks and decoding attacks. 

\subsection{Direct Attack}

A direct attack can be implemented either by finding vector $\bf u$ or $\bf e$ for a given ${\bf c}={\bf u}{\bf G}'+{\bf e}{\bf G}'_2$, ${\bf G}'$ and ${\bf G}'_2$. It takes at most $q^k$ rounds to recover ${\bf u}$ directly from $\bf c$. Since rank of ${\bf G}'_2$ is $n-k$ then $q^{n-k}$ rounds are required to recover $\bf e$ from $\bf c$ even when weight of $\bf e$ is an  arbitrary. To recover $\bf u$ from $\bf e$ an attacker has to solve the equation ${\bf  c}-{\bf e}{\bf G}'_2={\bf u}{\bf G}'$. Let us show that this equation has unique solution $\bf e$ such  that ${\bf  c}-{\bf e}{\bf G}'_2\in C$.
Let us assume that there are two error vectors ${\bf e}_1,{\bf e}_2$ that corresponds to the same codeword. It means that 
\begin{equation*}
 \begin{cases}
   {\bf c}-{\bf e}_1{\bf G}'_2={\bf u}{\bf G}' 
   \\
   {\bf c}-{\bf e}_2{\bf G}'_2={\bf u}{\bf G}' 
 \end{cases}
\end{equation*}
or
$$
\left({\bf e}_1-{\bf e}_2\right){\bf Q}({\bf G}_0+{\bf T})={\bf 0},
$$
since
$\left({\bf e}_1-{\bf e}_2\right){\bf Q}{\bf G}_0\in C$, then the number of solutions of $\left({\bf e}_1-{\bf e}_2\right){\bf Q}({\bf G}_0+{\bf T})={\bf 0}$ equals to the number of codewords of $C$ that can be represented as $\left({\bf e}_1-{\bf e}_2\right){\bf QT}$. Let us assume that ${\bf c}'\in C$ and ${\bf c}'=\left({\bf e}_1-{\bf e}_2\right){\bf QT}$. Since ${\bf QT}_{\mathcal{J}}={\bf 0}$, where $\mathcal{J}$ is an information set of $C$, then ${\bf c}'_{\mathcal{J}}={\bf 0}$ for any ${\bf e}_1$ and ${\bf e}_2$. It means  that ${\bf c}'={\bf 0}$ and thus $\left({\bf e}_1-{\bf e}_2\right){\bf Q}({\bf G}_0+{\bf T})={\bf 0}$ has unique solution and ${\bf e}_1={\bf e}_2$.

Moreover, using previous reasoning it can be proved that the system
\begin{equation*}
 \begin{cases}
   {\bf c}-{\bf e}_1{\bf G}'_2={\bf u}_1{\bf G}' 
   \\
   {\bf c}-{\bf e}_2{\bf G}'_2={\bf u}_2{\bf G}' 
 \end{cases}
\end{equation*}
can be fulfilled only when ${\bf u}_1={\bf u}_2$ and ${\bf e}_1={\bf e}_2$.

Summarizing all we can conclude that complexity of direct attack is at most $q^{\min\{k,n-k\}}$.

\subsection{Direct Decoding Attack}

The main idea of direct decoding attack is to find the unknown error vector $\bf e$ from the known ciphertext $\bf c$ and known ${\bf G}'={\bf G}{\bf M}$. Since ciphertext has the form ${\bf c}={\bf u}{\bf G}'+{\bf e}{\bf G}'_2$, then intruder has to decode in some coset ${\bf e}{\bf QT}+C$ of code $C$ where the weight of error vector ${\bf e}{\bf QTM}$ is up to $n-k$. But even if this error vector can be found, the obtained code word will have the form ${\bf u}{\bf G}{\bf M}+{\bf e}{\bf Q}{\bf G}_0{\bf M}$ and it is  difficult to recover $\bf u$ without knowledge of $\bf M$, $\bf G$, ${\bf G}_0$, $\bf T$ and $\bf Q$. Thus any direct decoding methods including ISD decoding can not been implemented to break the cryptosystem.

\subsection{Syndrome-Based Decoding Attack}

In this section we discuss sydrome-based attack. We are going to show that syndrome-based attack is infeasible to the proposed cryptosystem. 

Let us remind how to implement syndrome-based attack. To use it an attacker has to contruct parity-check matrix ${\bf H}'$ for known matrix ${\bf G}'={\bf G}{\bf M}$. It is easy to check that
$$
{\bf H}'={\bf L}{\bf H}\left({\bf M}^{-1}\right)^{T},
$$
where $\bf H$ is a parity-check matrix, corresponds to generating matrix $\bf G$ and $\bf L$ is some non-singular $(n-k)\times(n-k)$ matrix over $\mathbb{F}_q$. It is clear that matrix $\bf L$ does not affect code properties. Thus we can assume that ${\bf L}={\bf I}$. In this case the syndrome $\bf S$ of ciphertext $\bf c$ is as follows:
$$
{\bf S}={\bf c}{\bf H}'^T={\bf e}{\bf Q}{\bf T}{\bf H}^T={\bf e}\left({\bf H}\left({\bf QT}\right)^T\right)^T.
$$
The latest equation means that intruder has to solve syndrome equation for parity-check matrix ${\bf H}\left({\bf QT}\right)^T$ where matrix ${\bf QT}$ is singular. In this case we can not state that the properties of code $C'$ with this parity-check matrix are the same as for code $C$ with parity-check matrix $\bf H$.
Moreover, next example shows that transformation $\left({\bf QT}\right)^T$ usually results in random code.

\begin{example}
\it
Let us consider $(63,24)$ BCH code. This code has minimum distance $15$ and can correct up to $7$ independent errors. Any information set of this code consists of $k=24$ elements and thus for the case when given information set is free of errors, then at most $n-k=29$ errors placed out of information set can be corrected. If we consider generating matrix of this code in canonical form, then we can assume that $J=\{1,2,\ldots,24\}$ is information set of the code. We randomly constructed $100$ different matrices $\bf QT$ of size $63\times 63$ such that $\left({\bf QT}\right)_J={\bf 0}$ and other $29$ columns forms $63\times 29$ full rank matrix. Fig. \ref{fig1} represents weight spectrum of codes with parity-check matrices in the form ${\bf H}\left({\bf QT}\right)^T$, where $\bf H$ is a parity-check matrix of $(63,24)$ BCH code (blue curve).

\begin{figure}[h!]
\center{\includegraphics[width=\columnwidth]{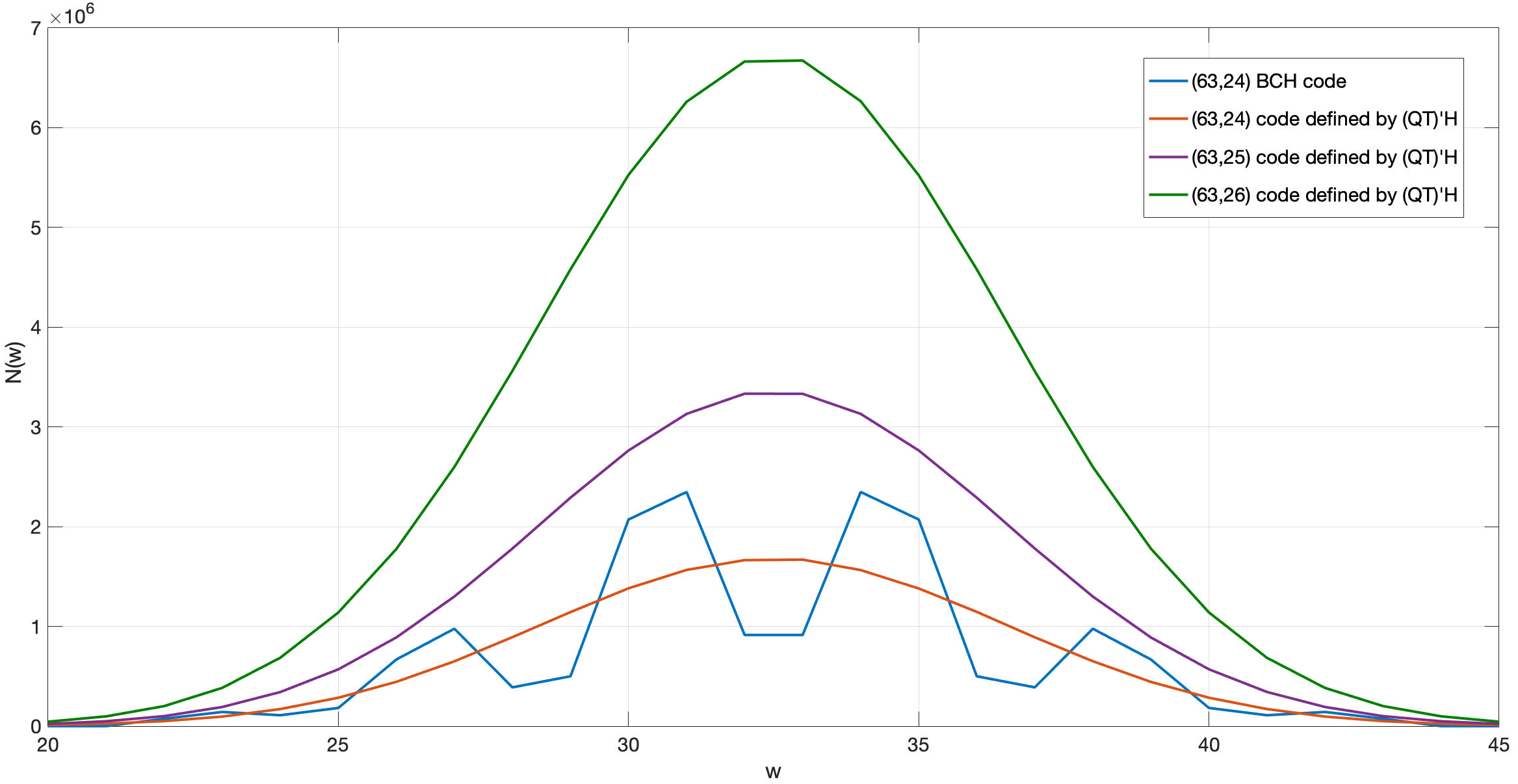} }
\caption{Spectra of codes defined by different parity-check matrices ${\bf H}\left({\bf QT}\right)^T$ obtained from $\bf H$ of $(63,24)$ BCH code}
\label{fig1}
\end{figure}

From Fig.~\ref{fig1} it can be noticed that singular transformation $\left({\bf QT}\right)^T$ does not hold code properties: even when code rate does not change, new code looks like random one with normal weight distribution (red curve). But there are some cases when linear transformation $\left({\bf QT}\right)^T$ leads to zero rows in ${\bf H}\left({\bf QT}\right)^T$: purple and green curves.  Purple curve corresponds to one zero row in ${\bf H}\left({\bf QT}\right)^T$ and thus corresponding code has parameters $(63,25)$ and green curve corresponds to two zero rows in ${\bf H}\left({\bf QT}\right)^T$ and the corresponding code has parameters $(63,26)$. Both $(63,25)$ and $(63,26)$ codes have normal weight distributions and thus can be considered  as random ones. 

Table 2 represents minimal distance properties of obtained codes.

\begin{center}
\setlength\tabcolsep{5pt}
\begin{table}[h!]
\label{tab1}
\caption{Minimal distance properties of 100 codes obtained from $(63,24)$ BCH code}
\begin{tabular}{|l|l|l|l|l|l|l|}
\hline
Code                     & \#                  & \multicolumn{5}{c|}{minimal distance}     \\ \hline
\multicolumn{2}{|l|}{} & min & max & average & variance & VG bound \\ \hline
(63,24)                  & 29                  & 10  & 13  & 11,28   & 0,56     & 9,66     \\ \hline
(63,25)                  & 60                  & 8   & 12  & 10,67   & 0,56     & 9,26     \\ \hline
(63,26)                  & 11                  & 9   & 11  & 10,18   & 0,56     & 8,87     \\ \hline
\end{tabular}
\end{table}
\end{center}

As it can be noticed transformation $\left({\bf QT}\right)^T$ usually results in one zero row in matrix ${\bf H}\left({\bf QT}\right)^T$ - the number of $(63,25)$ codes is $60$ and thus the probability to obtain code with this parameters from $(63,24)$ BCH code is approximately $0.6$. Moreover, the variance of minimal distance in all cases is rather small and equals to $\approx 0,56$. In the last column of the table the minimal distances of $(63,24)$, $(63,25)$, $(63,26)$ codes on Varshamov-Gilbert (VG) bound are represented. As it can be noticed, the obtained average values of minimal distances are rather close to the ones on VG bound. Thus the transformation $\left({\bf QT}\right)^T$ usually results in random code. \end{example}

The previous example and reasoning shows that sydrome-based attack is infeasible for the proposed cryptosystem. There are at least two reasons: parity-check matrix ${\bf H}\left({\bf QT}\right)^T$ is random, thus we can always choose matrix $\bf H$ that allows to correct significantly more errors than random code. The second reason is that the weight of $\bf e$ is arbitrary and thus it is not a coset leader. It means that applying any decoding algorithm an intruder usually find such ${\bf e}'$: $d({\bf c}-{\bf e}'{\bf G}_2',{\bf u}{\bf G}')<d({\bf c}-{\bf e}{\bf G}_2',{\bf u}{\bf G}')$, where $\bf e$ is valid error vector.

\section{Open key size}
In the first point of view the modification of McEliece cryptosystem can only increase an open key size since in this case length of the key is $n^2+k(n-k)$ while the open key original McEliece cryptosystem has length $k(n-k)$. But it was shown  that the best-known attack based on Information Set Decoding (ISD) can not be applied for this modification and thus this cryptosystem can be based on codes with significantly smaller lengths than of ones for ordinary McEliece cryptosystem.

Let us consider the following examples:
\begin{example}
\it
The initial parameters of McEliece cryptosystem are as follows: $n=1024$, $k=524$, $t=50$, thus $\log_2{\tau}\approx53$.  In order to obtain the same workfactor as for original McEliece cryptosystem in modified cryptosystem we need to find such code $C'$ with length $n'$ and correcting performance $k'$ that $\min\{k',n'-k'\}\approx53$. 

BCH Code with parameters $(n',k',t')=(127,71,9)$ has workfactor $\log_2{\tau'}\approx56$. In the case of $(1024,524,50)$ size of the open key is $k(n-k)=262000$ bits and in proposed cryptosystem nearly the same workfactor can be obtained for open key size $n'^2+k'(n'-k')=20105$ bits which is more than $13$ times smaller.
\end{example}

\begin{example}
\it 
To obtain workfactor $80$ the following parameters of original McEliece cryprosystem are used: $n=2048, k=1751, t=27$. This results in key size of approximately $520000$ bits. 

At the same time in modified cryptosystem the based on BCH code with parameters $n'=255,k'=79,t'=27$ has workfactor $79$ and open key size $78929$ bits which is approximately $6.6$ times smaller. 
\end{example}

\begin{example}
\it 
For stability against a quantum computer, the parameters should be increased to $n = 6960, k = 5413, t = 119$, and the size of the public key grows is up to $8373911$ bits. The workfactor of such cryptosystem will be approximately $263$. 

At the same time in modified cryptosystem the based on BCH code with parameters $n'=1023,k'=268,t'=103$ has workfactor $268$ and open key size $1818894$ bits which is approximately $4.6$ times smaller. 
\end{example}

All these examples show that proposed cryptosystem allows to significantly reduce open key size. Moreover, it does not depend on code which underlines McEliece cryptosystem. Thus this modification can be used jointly with code substitution, providing further reduction in public key size. 

\section{Conclusion}

In this paper we considered a new construction of well-known McEliece cryptosystem that allows to reduce the public key size and can be implemented for an arbitrary class of codes, i. e. it is code-independent. We considered some significant limitations on the parts of this cryptosystem and explained why best-known attacks against McElice cryptosystem based on information set decoding can not be implemented for the proposed modification. We also suggested some examples which  demonstrated that open key size can be reduced in several times.

\end{document}